\documentclass{article}
\usepackage[utf8]{inputenc}
\usepackage{kotex}

\title{Defining C-ITS Environment and Attack Scenarios}
\author{고려대학교 정보보호대학원 해킹대응기술연구실 \\ (김용식, 최재웅, 이효선, 유정도, 김해린, 장준호, 박기범, 김휘강)}
\date{November 2022}

\usepackage[numbers]{natbib}
\usepackage{graphicx}
\usepackage{amsmath}
\usepackage{amssymb}
\usepackage{verbatim}
\usepackage{cleveref}
\usepackage{subfig}
\usepackage{titlesec}
\usepackage{makecell}
\usepackage{multirow}
\usepackage{multicol}
\usepackage{booktabs}
\usepackage{adjustbox}

\titleformat{\paragraph}
{\normalfont\normalsize\bfseries}{\theparagraph}{1em}{}
\titlespacing*{\paragraph}
{0pt}{3.25ex plus 1ex minus .2ex}{1.5ex plus .2ex}

\begin{document}

\maketitle

\begin{abstract}
기술이 발전하면서 많은 데이터를 처리할 수 있게 되고, 도시 내 여러 요소가 다양화되고 복잡해지면서 도시들이 스마트 시티화 되고 있다.
스마트 시티의 핵심 시스템 중 하나로 Cooperative-Intelligent Transport Systems (C-ITS)가 있다.
C-ITS는 차량이 주행 중 운전자에게 주변 교통상황과 급정거, 낙하물 등의 사고 위험 정보를 노변 기지국을 통하여 실시간으로 제공하는 시스템이며, 도로 인프라, C-ITS 센터, 차량 단말기로 구성된다.
한편, 스마트 시티는 도시의 많은 요소들이 네트워크로 연결되고 전자적으로 제어되기 때문에 사이버 보안 문제가 발생할 수 있다.
C-ITS에서 사이버 보안 문제가 발생할 경우에는 안전 문제가 발생할 위험이 크다.
본 기술문서는 C-ITS 보안을 위한 C-ITS 환경 모델링과 C-ITS 공격 시나리오를 기술하는 것을 목적으로 한다.
C-ITS의 개념과 MITRE ATT\&CK에 대해 기술한 뒤, 우리가 정의한 C-ITS 환경 모델과 공격 시나리오 모델을 기술한다.
\end{abstract}


\begin{center}
\subsubsection*{Acknowledgment}
This work was supported by Institute of Information \& Communications Technology Planning \& Evaluation (IITP) grant funded by the Korea government (MSIT) (No. 2021-0-00624, Development of Intelligence Cyber Attack and Defense Analysis Framework for Increasing Security Level of C-ITS)
\end{center}

\section{인트로}
스마트 시티는 다양한 데이터를 수집하고, 이 데이터를 도시 운영에 사용하는 도시이다.
기술이 발전하면서 많은 데이터를 처리할 수 있게 되고, 도시 내 여러 요소가 다양화되고 복잡해지면서 도시들이 스마트 시티화 되고 있다.
스마트 시티의 핵심 시스템 중 하나로 Cooperative-Intelligent Transport Systems (C-ITS)가 있다.
C-ITS는 교통체계를 제어하며, 이를 위해 도시 내 차량과 교통 인프라의 정보를 이용한다.
도시에서의 인적, 물적 교류는 매우 중요한 문제이기 때문에 이를 효율적으로 제어할 수 있는 C-ITS의 필요성은 매우 높다.

한편, 스마트 시티는 도시의 많은 요소들이 네트워크로 연결되고 전자적으로 제어되기 때문에 사이버 보안 문제가 발생할 수 있다.
C-ITS의 차량이나 교통 인프라도 예외는 아니다.
C-ITS에서 사이버 보안 문제가 발생할 경우에는 안전 문제가 발생할 위험이 크다.
차량과 교통 인프라가 잘못 제어된다면 도시 내 인적 물적 교류가 막히고, 심각한 경우에는 차량 사고가 발생할 수도 있다.
사이버 보안 문제를 예방하고 최소화하기 위해서는 C-ITS 보안 대책이 필요하다.

본 기술문서는 C-ITS 보안을 위한 C-ITS 환경 모델링과 C-ITS 공격 시나리오를 기술하는 것을 목적으로 한다.
Section \ref{background}에서는 C-ITS의 개념과 MITRE ATT\&CK에 대해 기술한다.
Section \ref{c-its-modeling}과 Section \ref{attack-scenario-modeling}에서는 각각 우리가 정의한 C-ITS 환경 모델과 공격 시나리오 모델을 기술한다.

\section{배경}
\label{background}
\subsection{C-ITS}
\subsubsection{C-ITS 개념}
C-ITS는 차량이 주행 중 운전자에게 주변 교통상황과 급정거, 낙하물 등의 사고 위험 정보를 노변 기지국을 통하여 실시간으로 제공하는 시스템이며, 도로 인프라, C-ITS 센터, 차량 단말기로 구성된다 \cite{cits}.

\subsubsection{C-ITS 필요성}
\paragraph{교통사고 예방을 통한 안전성과 이동성 향상}
C-ITS는 Vehicle-to-Vehicle (V2V), Vehicle-to-Infrastructure (V2I) 통신 기반의 차량, 인프라간의 정보 공유와 C-ITS 센터와 노변 기지국을 활용한 실시간 정보 수집$\cdot$제공$\cdot$연계$\cdot$위치 기반 서비스를 사용자에게 제공하여, 교통 사고를 예방하고 안전성과 이동성을 향상 시킨다.
\paragraph{도로관리 중심에서 이용자안전 중심의 패러다임 변화}
C-ITS의 등장으로 교통소통정보, 가공된 도로 교통정보 위주였던 Intelligent Transport Systems (ITS)에서 교통안전정보, 실시간 정보를 제공하는 교통 흐름 패러다임의 변화를 가져왔다.
또한, 실시간 정보 및 교통안전정보를 바탕으로 즉시대응의 한계점이 명확했던 기존 ITS에서 벗어나, 사후관리 중점에서 사전에 사고를 예방하는 이용자 안전 중심의 패러다임 변화를 가져왔다.

\subsubsection{자율주행차량과 C-ITS 간의 관계}
\paragraph{자율주행차량의 한계 극복을 위한 도로 인프라 지원}
자율주행차량의 경우 악천우시 차량센서의 기능이 저하 된다는 점과 원거리 감지 기능 및 사각지대 검지 한계점을 극복하기 위해 노변 기지국을 활용한 실시간 정보를 제공한다.
\paragraph{C-AV}
C-AV란, CV(Connected Vehicle)와 AV(Autonomous Vehicle)를 합친 개념으로, 네트워크와 연결되어 각종 정보를 제공 받는 개념의 CV와 자율적으로 차량을 운전하는 AV의 개념을 합친 C-AV의 필요성을 인지하고 자율주행 시대를 대비해 C-ITS의 역할이 중요해지고 있다.
\subsubsection{C-ITS 제공 서비스}
C-ITS가 제공하는 서비스는 다음과 같다.
\begin{multicols}{2}
    \begin{itemize}
        \item 위치기반 데이터 수집
        \item 위치기반 교통정보 제공
        \item 요금징수시스템
        \item 노면 기상정보 제공
        \item 도로위험구간 정보제공
        \item 보행자 충돌방지 경고
        \item 차량 긴급상황 경고
    \end{itemize}
\end{multicols}

\subsubsection{C-ITS 통신}
\begin{figure}[!ht]
    \centering
    \includegraphics[width=6cm,height=6cm]{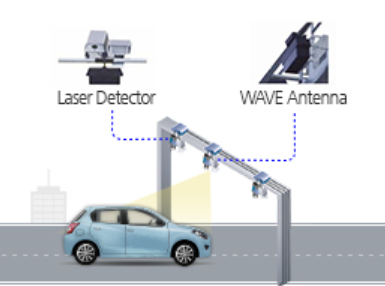}
    \caption{WAVE protocol.}
    \label{fig_WAVE_networking}
\end{figure}

\paragraph{개요}
\begin{figure}[ht!]
    \centering
    \includegraphics[width=13cm,height=8cm]{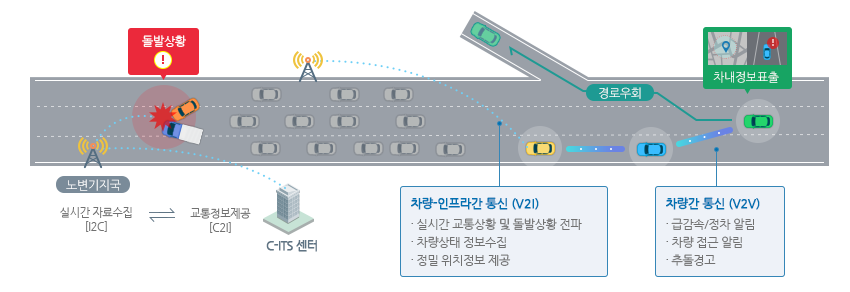}
    \caption{C-ITS Infrastructure.}
    \label{fig_cits}
\end{figure}
C-ITS 통신에서는 차량,노드 간의 통신을 위해 Wireless Access in Vehicular Environments(WAVE)를 사용한다. WAVE는 WLAN (IEEE 802.11a, Wi-Fi) 기술을 기반으로 자동차 운용 환경에 맞도록 개선하여 고속으로 주행하는 차량에 통신 서비스를 제공하는 것에 특화된 차세대 통신 기술이다. WAVE의 성능은 최대 200km/h로 주행하는 차량과 통신이 가능하며, 최대 1km까지 12$\sim$27Mbps 전송 속도로 100m/sec 이내에 전파 전달이 가능하다. 현재 하이 패스 등에 안전한 메세지 통신을 위해 사용되는 단거리 전용 통신(Dedicated Short-Range Communication, DSRC)의 일종이며,노변 기지국과 같은 인프라와 차량간의 통신을 위한 Vehicle-to-Infrastructure (V2I)$\cdot$차량간의 통신을 위한 Vehicle-to-Vehicle (V2V) 통신을 지원한다. C-ITS 통신의 구조는 \cref{fig_cits}과 같으며, WAVE 통신 프로토콜은 \cref{fig_WAVE_networking}의 구조로 동작한다.

\begin{figure}[ht!]
\centering
\subfloat[V2V on FV]{
\includegraphics[width=0.35\linewidth]{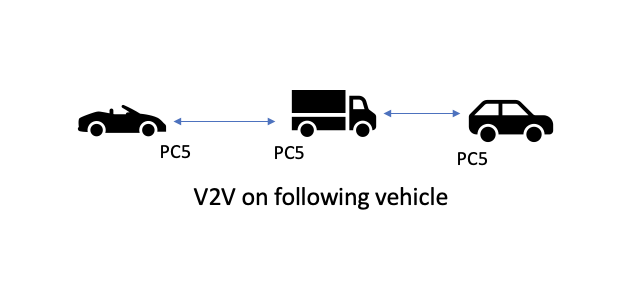}
\label{LTEV2V}
}
\centering
\subfloat[V2I on FV\&Uu]{
\includegraphics[width=0.35\linewidth]{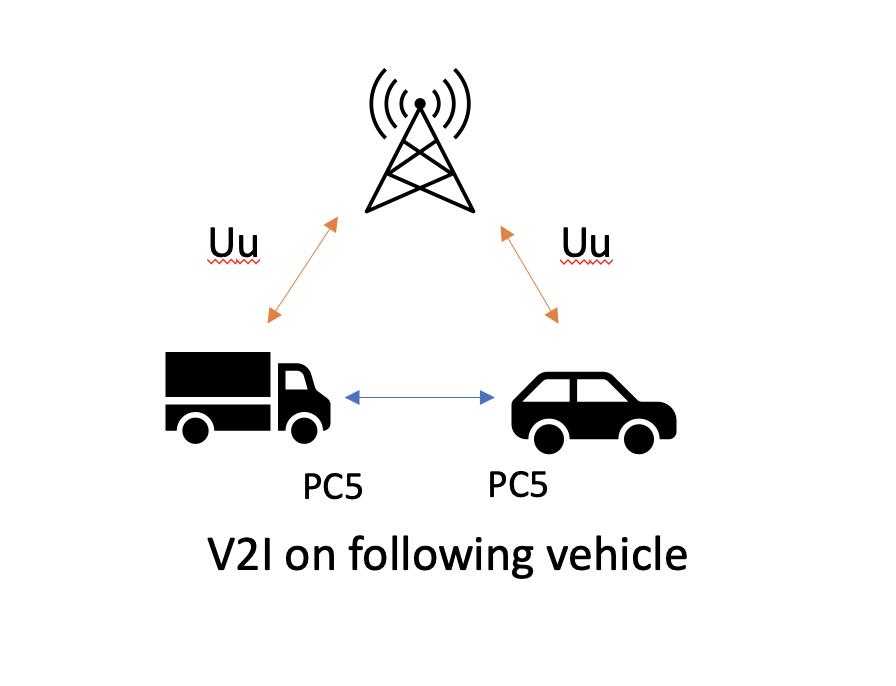}
\label{LTEV2I}
}
\caption{LTE V2X Communication Example}
\label{LTE V2X Communication Example}
\end{figure}

\paragraph{LTE V2X}
국내에서는 기존의 LTE 통신 기술을 차량통신에 적합하도록 자원할당 방식을 개선해 사용하는 C-V2X 기술을 이용하여 C-ITS 환경에서 차량과 차량간의 통신, 차량과 기지국 간의 통신을 지원한다. \cref{LTEV2V}와 같이 선행 차량과 후속 차량 간의 별도의 인터페이스 PC5를 이용해 통신할 수 있고, \cref{LTEV2I}와 같이 차량과 기지국 간의 통신(V2I)을 위해 기지국 인터페이스 Uu로 정보를 전달할 수 있다. 
WAVE 방식과의 차이점은 이미 ISP 측에서 구축한 LTE 기지국을 사용하기 때문에 별도의 기지국을 구축할 필요가 없고, 대역폭 또한 최대 75Mbps로 27Mbps인 WAVE보다 빠르기 때문에 정보 제공 뿐만 아니라 동영상과 같은 미디어 콘텐츠 소비 또한 latency 없이 사용할 수 있다.
LTEX V2X는 군집 주행, 차량 센서와 V2X 통신을 이용한 협력 인지 원격 제어 서비스 등의 자율 주행을 위한 통신을 포함하고 있다. 자율 주행을 위한 무선 통신기술은 latency와 신뢰성이 중요하며 10m/sec 정도의 Latency와 99.999\% 정도의 신뢰성을 목표로 하고 있다. 이러한 성능은 LTE V2X에서 제시하고 있는 물리 계층의 프레임 구조와 MAC 계층에서의 무선 자원 제어 기술에 의해 직접적으로 영향을 받게 되므로 응용 서비스에 따라 요구하는 Quality of Service 를 만족하기 위한 기술과 주파수 영역에서의 자원 관리 기술이 핵심기술이 될 것이다.

\begin{figure}[ht]
    \centering
    \includegraphics[width=6cm,height=4cm]{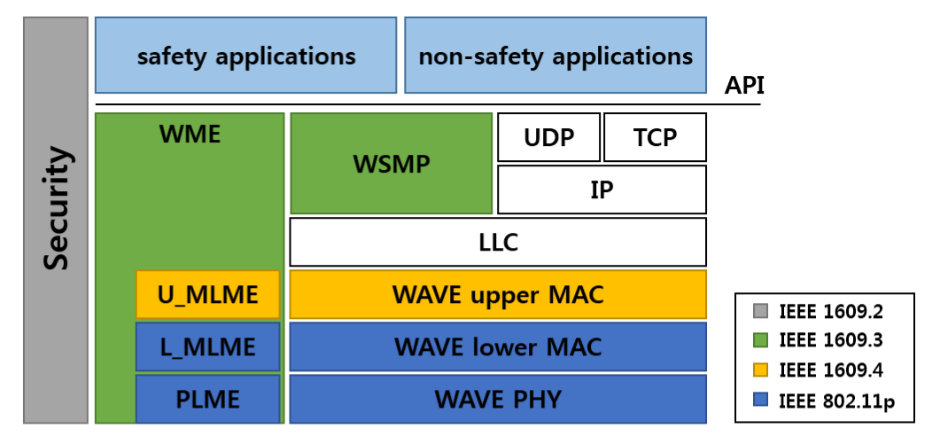}
    \caption{IEEE 1609.2 Structure.}
    \label{fig_IEEE1609.2}
\end{figure}

\paragraph{IEEE 1609.2} 
IEEE 1609.2는 C-ITS 및 자율협력주행 기술에 적용되는 Vehicle-to-Everything(V2X) 통신의 보안성을 확보하는 표준이다. WAVE 통신 관련 보안 표준으로, 메시지 위$\cdot$변조 방지, 전자서명 기반 송신자 검증, AES-CCM 기반 메시지 암호화를 제공한다. 또한, MAC 값을 추가적으로 생성하기 때문에 데이터의 무결성이 보장된다. IEEE 1609.2 표준의 구조는 \cref{fig_IEEE1609.2}와 같다.
\paragraph{IEEE 802.11p}
IEEE 802.11p는 차량간 무선 접속을 통한 통신이나 차량과 노변 기지국과의 WAVE 통신을 위한 물리계층과 MAC 계층을 정의한다.

\subsection{MITRE ATT\&CK}
\subsubsection{사이버 킬체인}
\begin{figure}[ht!]
    \centering
    \includegraphics[width=8cm,height=8cm]{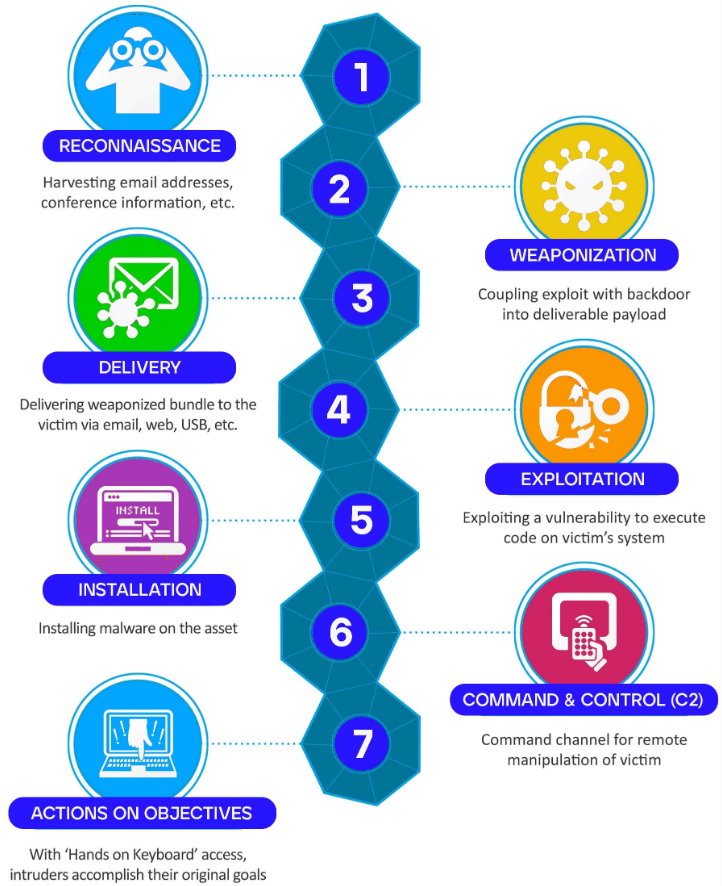}
    \caption{Cyber Killchain overview.}
    \label{fig_cyberkillchain}
\end{figure}

사이버 킬체인이란, 악의적인 사이버 공격을 단계별로 분석하여 대상에게 가해지는 위협 요소를 파악하고, 공격자의 의도를 완화시키는 전략을 의미한다. 사이버 킬체인은 모든 공격을 다 막아낼 수 없기 때문에 공격자 입장에서의 공격 분석을 통해 공격 단계의 연결 고리를 끊어 피해를 최소화 하기 위한 방법이다. 사이버 킬체인은 \cref{fig_cyberkillchain}과 같은 구조로 구성되어 있으며, 기본적으로 해당 흐름을 따른다.
MITRE ATT\&CK과 마찬가지로 사이버 공격에 관련하여 기술한 데이터라는 점에서 유사성이 있다.
다만, 사이버 킬체인은 사이버 공격을 선형적으로 단계별 서술을 했을 뿐, 각 공격 단계와 연관된 공격 도구나 해킹그룹을 연결짓기 어렵다. 그리고 외부 침입을 막기 위한 모델이기 때문에 이미 발생한 침입이나 내부자에 의한 공격에 대한 전략이 미흡하다. 이 점을 보완하기 위하여 공격 행위를 Tactics와 Techniques의 관점으로 모델링한 것이 MITRE ATT\&CK이다.

\subsubsection{MITRE ATT\&CK 개요}
\begin{figure}[ht!]
    \centering
    \includegraphics[width=1\linewidth]{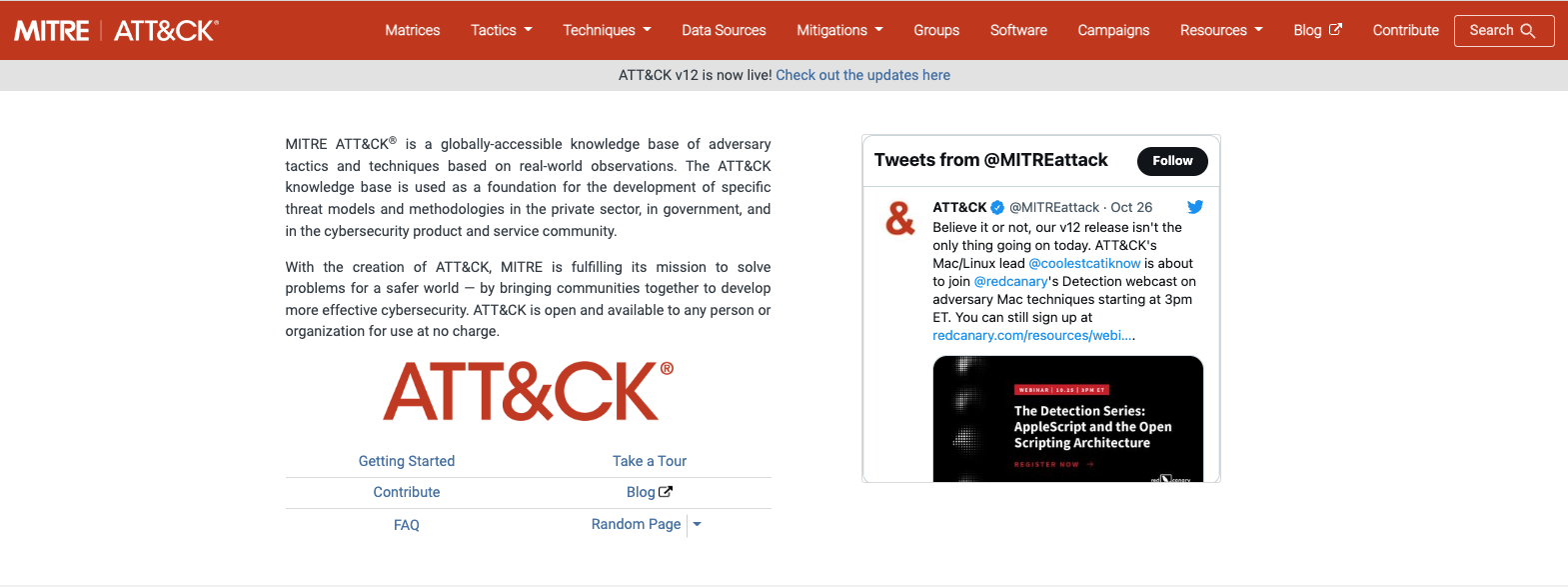}
    \caption{MITRE ATT\&CK.}
    \label{fig_MITRE_ATT_CK}
\end{figure}
MITRE ATT\&CK란, \cref{fig_MITRE_ATT_CK}와 같이 공격자들의 최신 공격 기술 정보가 담긴 저장소를 의미한다. MITRE ATT\&CK에서의 ATT\&CK는  Adversarial Tactics, Techniques, and Common Knowledge의 약어로 사이버 공격 사례를 관찰한 후 공격자가 사용한 악의적 행위에 대해 공격방법과 기술의 관점으로 분석하여 다양한 공격 그룹의 공격기법들에 대한 정보를 분류해 목록화 해 놓은 표준적인 데이터들이다. 
윈도우 기업 네트워크 환경에서 사용되는 해킹 공격에 대해 방법(Tactics), 기술(Techniques), 절차(Procedures) 등 TTPs를 문서화하는 것을 시작으로 공격자로부터 발생한 일관된 공격 행동 패턴에 대한 분석을 기반으로 TTPs 정보를 매핑하여 공격자의 행위를 식별해 줄 수 있는 프레임워크를 제공한다.

\subsubsection{MITRE ATT\&CK Framwork}
\begin{figure}[ht!]
    \centering
    \includegraphics[width=1\linewidth]{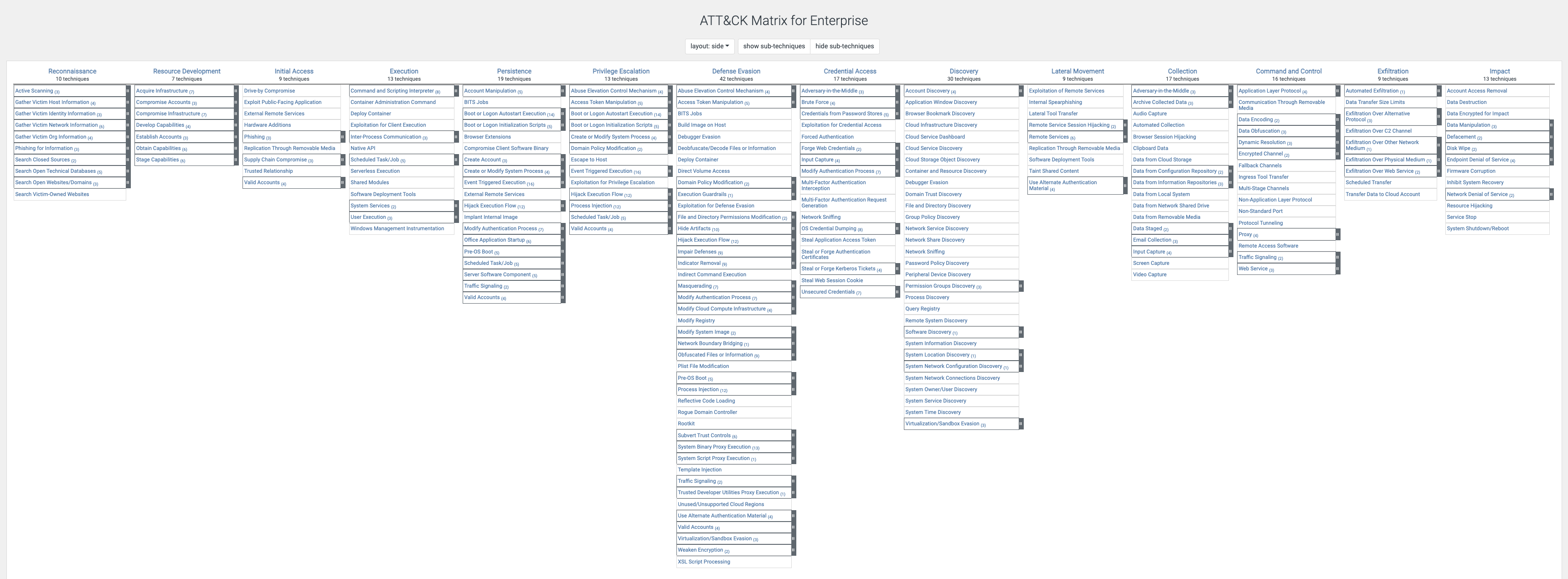}
    \caption{ATT\&CK Matrix for Enterprise.}
    \label{fig_ATT_CK_Mat_Ent}
\end{figure}

\paragraph{ATT\&CK Matrix}
Matrix란, \cref{fig_ATT_CK_Mat_Ent}와 같이 공격 기술인 Tactics, Techniques 개념과 관계를 시각화 한 것이다. 필요에 따라 Enterprise, Mobile, ICS 버전으로 제공한다. Enterprise 버전은 범용적인 기업환경에 적용되는 네트워크와 OS에 대해 기업 침해 행위를 세부적으로 모델링하기 위한 프레임워크이다. Mobile 버전은 모바일 OS에 대한 침해 행위를 세부적으로 모델링하기 위한 프레임워크이다. ICS 버전은 산업제어시스템에 산업 생산 영역에서 설비의 운영을 제어$\cdot$관리하는 시스템을 대상으로 한 공격 유형 정보를 포함하고 있다.

\begin{figure}[ht!]
    \centering
    \includegraphics[width=0.75\linewidth]{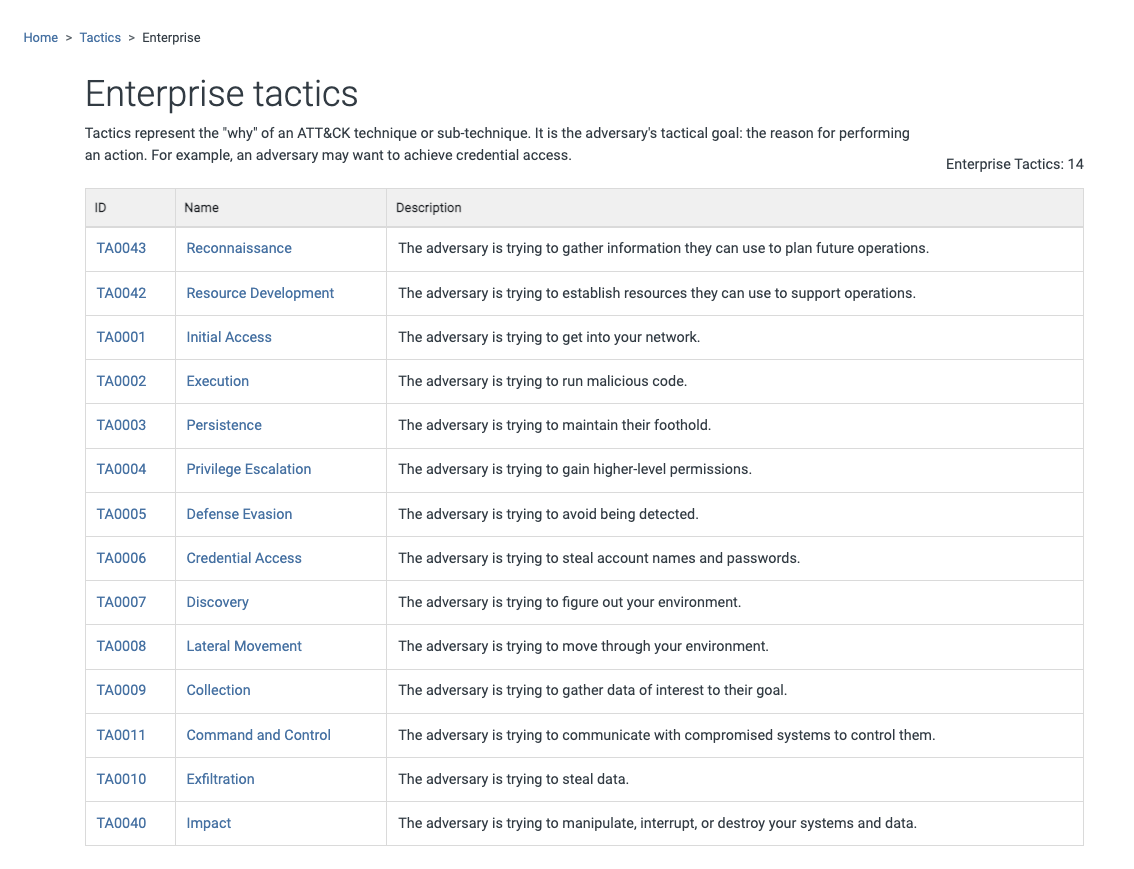}
    \caption{ATT\&CK Tactics for Enterprise.}
    \label{fig_ATT_CK_Tac}
\end{figure}

\paragraph{Tactics}
Tactics란, 공격자의 공격 목표에 따른 행동을 나타낸다. 상황별 각각의 Techniques에 따른 범주 역할이다. 이는 공격자의 전술적인 목적으로 행위를 수행하는 이유이다. 예를 들어, 공격자는 credential access를 달성하기를 원한다. 공격 목적에 따라 지속성, 정보탐색, 실행, 파일 추출 등 다양한게 분류되며 Enterprise 14개, Mobile 14개, ICS 12개로 구성된다.
Enterprise의 Tactics는 \cref{fig_ATT_CK_Tac}과 같으며, 14개 Tactics에 대해 ID와 이름, 설명으로 구성되어 있다. 

\begin{figure*}[ht]
\centering
\subfloat[Techniques]{
    \includegraphics[width=0.45\linewidth]{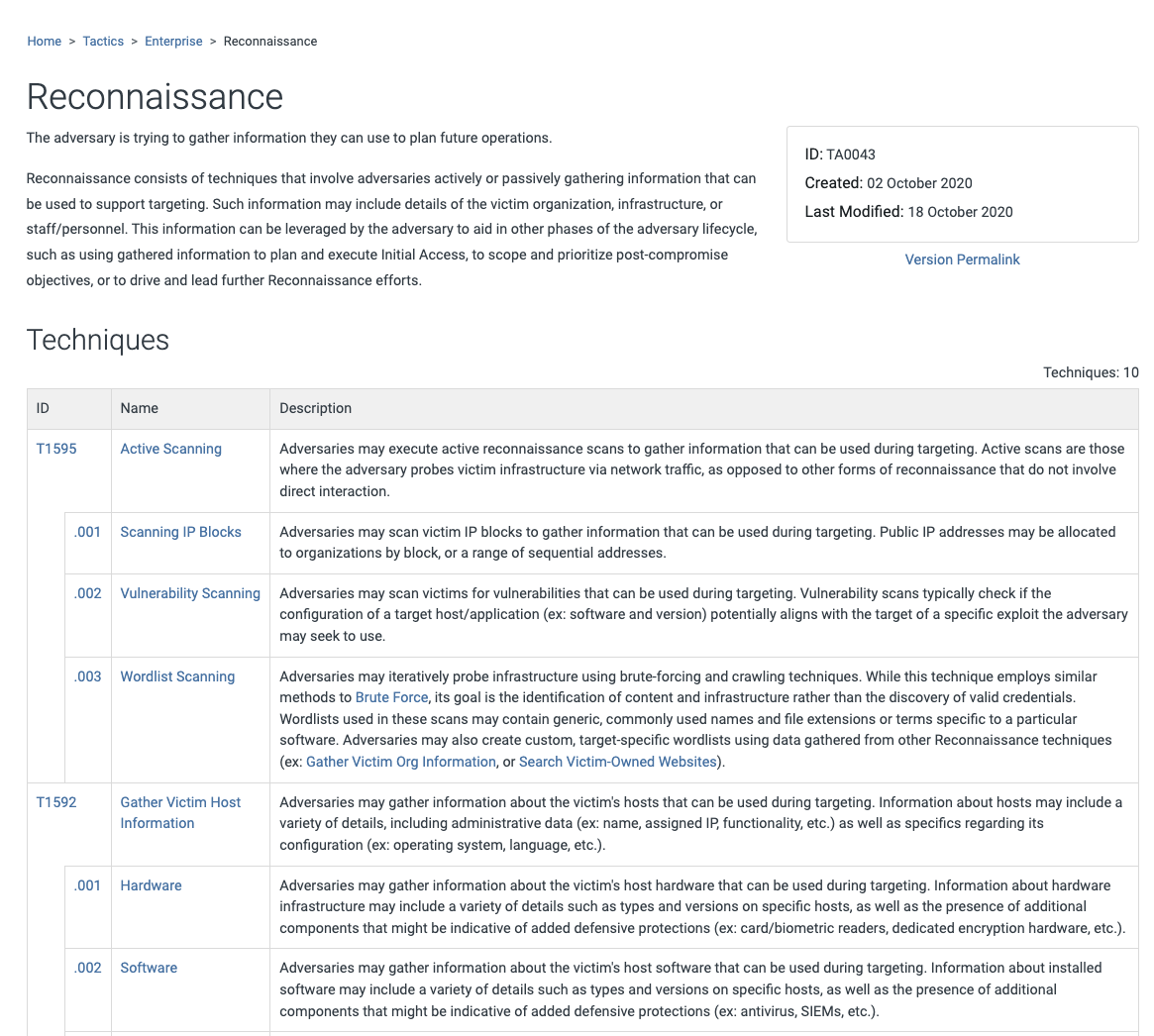}
    \label{fig_ATT_CK_Teh}
    }
\hfil
\subfloat[Active Scanning of Techniques]{
    \includegraphics[width=0.45\linewidth]{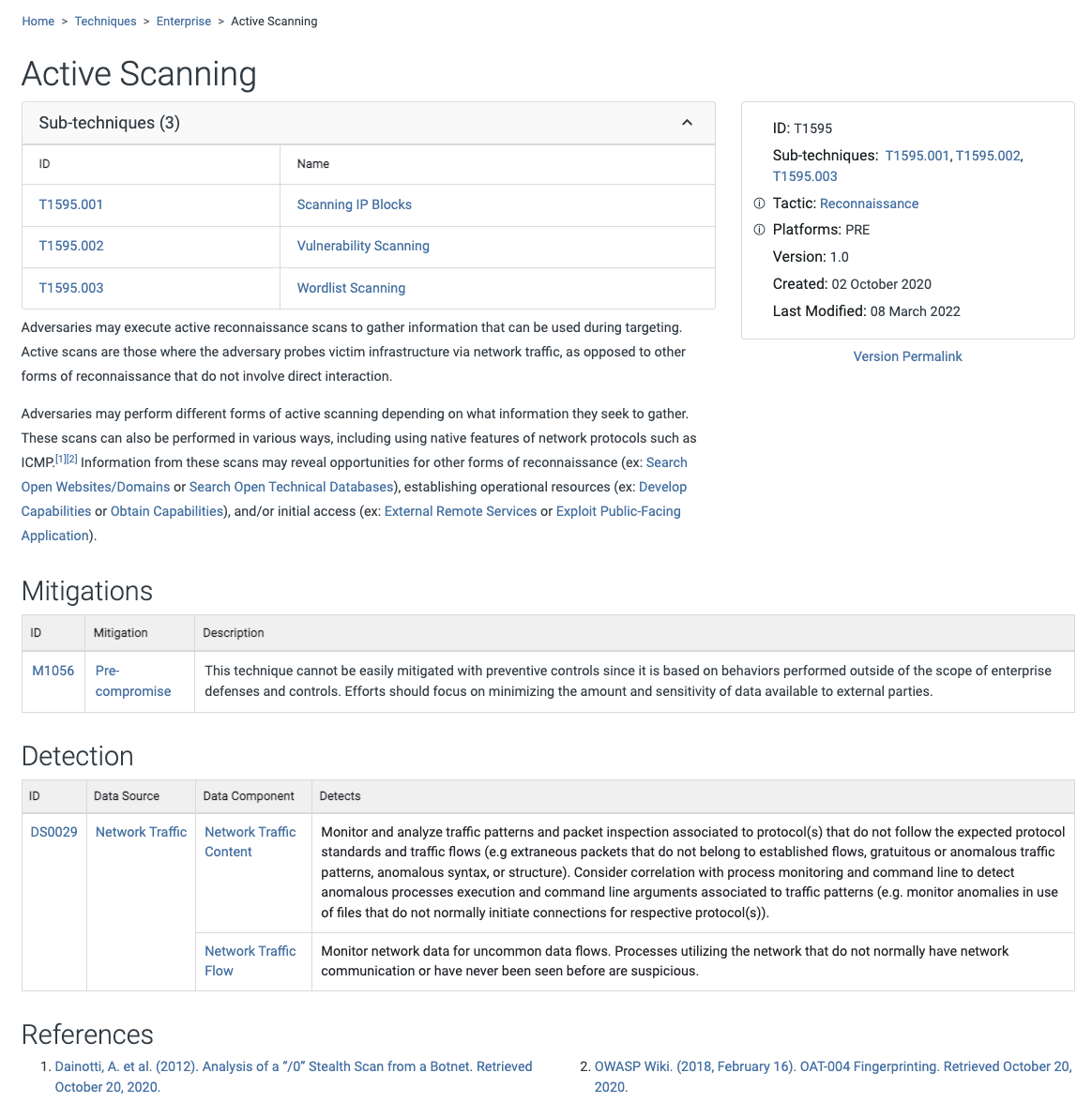}
    \label{fig_ATT_CK_AS}
    }
\caption{ATT\&CK Techniques for Enterprise.}
\label{fig_ATT_CK_T}
\end{figure*}

\paragraph{Techniques}
Techniques란, 공격자가 목표에 대한 Tactics를 달성하기 위한 방법을 나타낸다. 즉, Techniques를 통해 발생하는 결과를 명시한다. 
Techniques 정보를 통해 실제 공격에 사용된 세부 기술을 확인할 수 있다.
예를 들어, \cref{fig_ATT_CK_Teh}에서와 같이 Reconnaissance가 있다. 
공격자는 그들의 향후 계획을 위해 정보를 모으려고 한다. 
Reconnaissance는 공격자들이 적극적이거나 수동적으로 타겟팅을 하는데 필요한 정보들을 모으려고 하는데 수반되는 기술들로 구성되어 있다. Techniques는 ID, 이름, 설명으로 기술되어 있다.  
세부적으로, \cref{fig_ATT_CK_AS}과 같이 Active Scanning이라는 Techniques가 있다. 
해당 Techniques에는 3개의 sub-techniques가 있다. 
Techniques에 대한 Mitigations와 Detection에 대해서도 설명하고 있다. 

\paragraph{Mitigations}
Mitigations란, 방어자가 공격을 예방하고 탐지하기 위해 취할 수 있는 행동을 의미한다.
MITRE ATT\&CK의 Mitigations는 보안 개념과 공격자의 Techniques나 sub-technique이 성공적으로 수행되는 것을 막기 위한 기술의 범주로 설명하고 있다. Enterprise Mitigations는 43개로 ID, 이름, 설명으로 기술되어 있다. 

\paragraph{Groups}
Groups란, 공개적으로 명칭이 부여된 해킹단체에 대한 정보와 공격 기법을 분석하여 정리한 범주이다. 주로 사용된 공격 방법과 문서 등을 바탕으로 조직을 특정하여 정의한다. 공격에 사용된 Techniques와 Software 목록을 포함하고 있고, 그룹이 자주 사용하는 공격 형태를 제공한다.

\paragraph{Software}
Software란, 공격자가 목표 대상을 공격할 때 사용된 공격 코드 또는 운영체제에 포함된 도구나 Open Source S/W 등을 목록화하여 범주화한다. 

\section{C-ITS 환경 모델링}
\label{c-its-modeling}
\begin{figure}[ht]
    \centering
    \includegraphics[width=7cm,height=7cm]{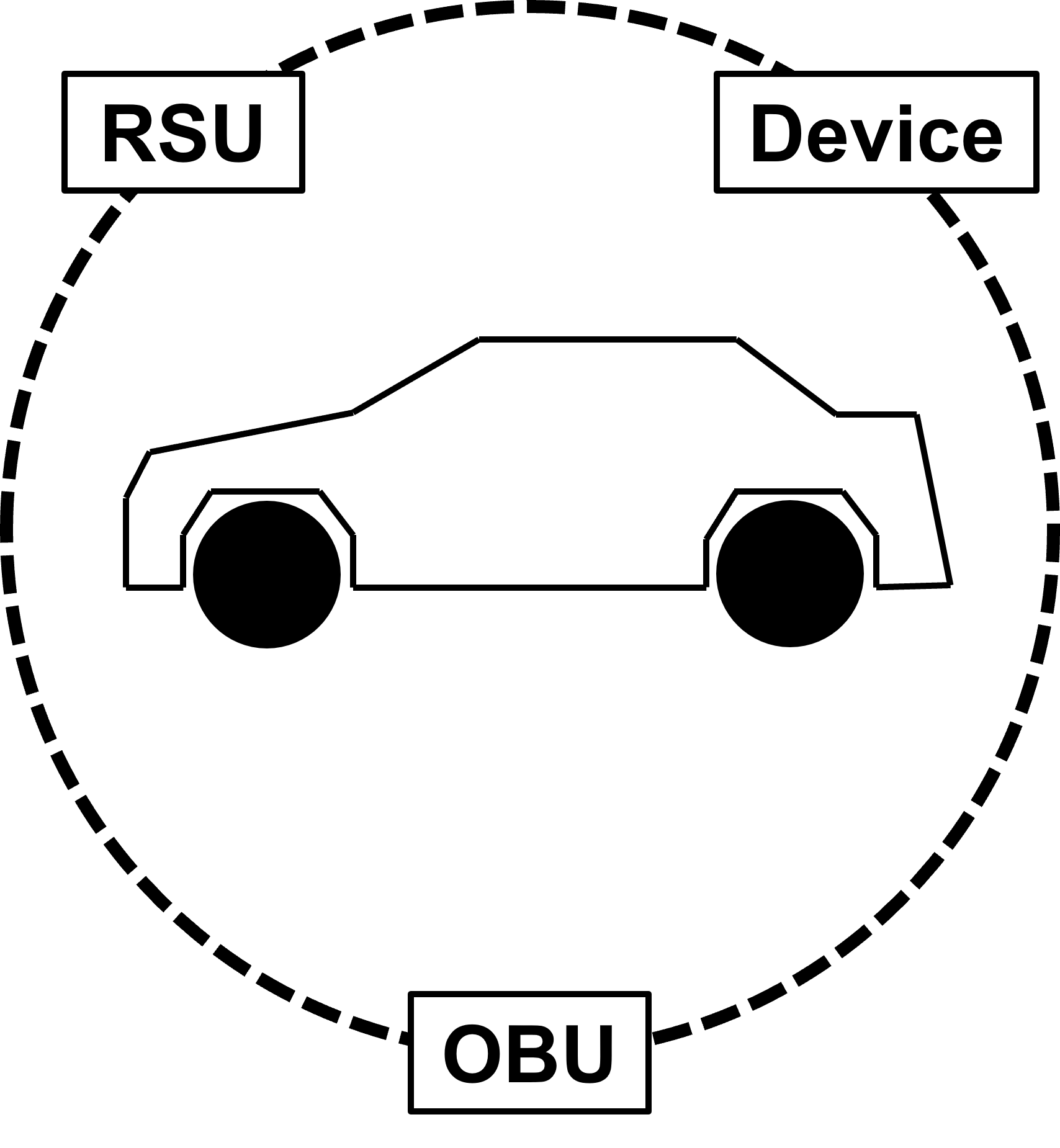}
    \caption{C-ITS environment modeling overview.}
    \label{fig_C-ITS_Env}
\end{figure}

\subsection{소개}
C-ITS 환경 모델링이란, \cref{fig_C-ITS_Env}와 같이, C-ITS 환경의 복합적인 구성요소 (OBU, RSU, 취약점, MITRE ATT\&CK 프레임워크, 네트워크 등)를 간결하게 표현한 것이다.
C-ITS 환경 기반 공격 그래프를 생성하기 위해서는 C-ITS 환경 데이터의 흐름을 파악할 수 있는 도식화가 요구되므로, C-ITS 환경 모델링을 수행해야 한다.
하지만, C-ITS 환경 모델링을 수행하기 위해서는 각 구성요소 및 구성요소 간 관계성을 정의해야 할 필요성이 존재한다.
본 장에서는 효과적인 C-ITS 환경 모델링을 수행하기 위해 각 환경 정보를 정의하였다.

\subsection{환경 정보}
\subsubsection{On Board Unit (OBU)}
OBU란, 차량 내에 설치되는 단말기를 의미하며, 구간 별 통행 속도$\cdot$위치정보 등을 수집하고, 해당 정보를 RSU를 거쳐 인터넷 망을 통해 C-ITS 센터로 전송한다.
환경 정보에서의 OBU는 도로 내 움직이는 차량 및 RSU와 통신할 수 있는 개체를 의미한다.
OBU는 다음과 같은 특성을 가진다.
\begin{itemize}
    \item OBU는 시간에 따라 가진 특성이 변화할 수 있다.
    \item OBU는 Road Side Unit (RSU)와 매 시간마다 정보를 송$\cdot$수신한다.
    \item OBU는 RSU에서 수신 받은 정보를 통해 경로를 변경할 수 있으나 최종 목적지는 변하지 않는다.
\end{itemize}
OBU 정보를 C-ITS 환경에서 구현하고, 활용하기 위해서는 다양한 정보를 간결한 데이터로 축약해야 하므로, 데이터 오브젝트를 키-값 쌍으로 구성할 수 있는 JSON 포맷을 사용하였다. OBU 정보를 표현하는 키 이름과 설명은 다음과 같이 표현된다.

\begin{itemize}
    \item id
        \begin{itemize}
            \item C-ITS 환경 내에서 각 OBU를 구분하는 고유 식별자를 의미하는 field를 의미한다.
        \end{itemize}
    \item Loc\_Road
        \begin{itemize}
            \item 현재 OBU가 위치한 도로 식별자를 의미하는 field를 의미한다.
        \end{itemize}
    \item Con\_Network
        \begin{itemize}
            \item 현재 OBU가 연결된 네트워크 식별자를 의미하는 field를 의미한다.
        \end{itemize}
    \item Description
        \begin{itemize}
            \item OBU의 상세 정보를 표현하는 field로, 각 OBU에 대한 설명 및 정보를 나타낸다. 
        \end{itemize}
    \item Destination
        \begin{itemize}
            \item OBU의 최종 목적지 (\textit{e.g.}, 내비게이션 목적지)를 나타내는 field로, 최종 attack step에서 OBU가 위치할 도로 식별자 field를 의미한다. 
        \end{itemize}
    \item System\_Env
        \begin{itemize}
            \item OBU가 가진 환경 정보 (\textit{e.g.}, 시스템) field를 의미한다.
        \end{itemize}
    \item Command
        \begin{itemize}
            \item OBU에서 발생된 Command를 의미하며, 내부에서 발생한 명령을 서술식으로 표현한다. 
        \end{itemize}
    \item Protocol
        \begin{itemize}
            \item OBU가 사용하는 Protocol 정보들을 표현하며, 본 C-ITS 환경 모델링의 경우 양 기기가 통신하기 위해서는 공통된 Protocol이 존재하여야 한다.
        \end{itemize}
\end{itemize}

또한, OBU 정보를 표현하는 데이터형과 그 예시는 \cref{des_OBU}과 같다.

\begin{table}[ht!]
    \begin{tabular}{c|l|c}
    \hline
    키 이름 & 데이터 타입 & 예시 \\\hline
    \hline
    id & String & OBU\_Vehicle\_hyundai\\\hline
    Loc\_Road & String & Road\_1 \\\hline
    Con\_Network & String & Network\_C-ITS \\\hline
    Description & String & Avante \\\hline
    Destination & String & Road\_2 \\\hline
    System\_Env & String Array &  [``Linux Yocto'', ``USB''] \\\hline
    Command & String Array & [``Receive mal-information from RSU''] \\\hline
    Protocol & String Array & [``Ethernet'', ``Bluetooth'', ``WTP'']\\\hline
    \end{tabular}
    \caption{Description of the On Board Unit.}
    \label{des_OBU}
\end{table}

\subsubsection{Road Side Unit (RSU)}
RSU란, 도로 주변 시설물에 설치되는 장비를 의미하며, C-ITS 센터에서 가공된 정보를 도로 내 OBU에 전달하는 역할을 수행한다.
환경정보에서의 RSU는 차량과 C-ITS 센터 사이 정보를 교환하는 노변기지국 (또는 장치)을 의미한다.
RSU는 다음과 같은 특성을 가진다.
\begin{itemize}
    \item RSU는 도로 교통 제어 및 차량이 운행에 필요한 정보를 전달한다.
    \item RSU는 OBU에서 얻은 정보를 C-ITS 센터로 송신하며, C-ITS 센터에서 수신 받은 명령 또는 정보를 OBU에 전달한다.
    \item RSU는 RSU와 통신을 통해 정보를 교환, 교환된 정보를 이용해 실시간으로 OBU 또는 C-ITS 센터에 송신한다.
\end{itemize}
RSU 정보 또한, 다양한 정보를 C-ITS 환경 모델링 상에서 표현하기 위해 JSON 포맷을 이용하여 키-값 쌍으로 구성하였다. RSU 정보를 표현하는 키 이름과 설명은 다음과 같이 표현된다.

\begin{itemize}
    \item id
        \begin{itemize}
            \item C-ITS 환경 내에서 각 RSU를 구분하는 고유 식별자를 의미하는 field를 의미한다.
        \end{itemize}
    \item Loc\_Road
        \begin{itemize}
            \item 현재 RSU가 위치한 도로 식별자를 의미하는 field를 의미한다.
        \end{itemize}
    \item Con\_Network
        \begin{itemize}
            \item 현재 RSU가 연결된 네트워크 식별자를 의미하는 field를 의미한다.
        \end{itemize}
    \item Groups
        \begin{itemize}
            \item RSU 기기가 속한 RSU 그룹을 의미하는 field로, 그룹 별 사용 목적이 다르므로 RSU는 그룹에 따라 사용 용도를 유추할 수 있다. 
        \end{itemize}
    \item Description
        \begin{itemize}
            \item RSU의 상세 정보를 표현하는 field로, 각 RSU에 대한 설명 및 정보를 나타낸다. 
        \end{itemize}
    \item System\_Env
        \begin{itemize}
            \item RSU가 가진 환경 정보 (\textit{e.g.}, 시스템) field를 의미한다.
        \end{itemize}
    \item Command
        \begin{itemize}
            \item RSU에서 발생된 Command를 의미하며, 내부에서 발생한 명령을 서술식으로 표현한다. 
        \end{itemize}
    \item Protocol
        \begin{itemize}
            \item RSU가 사용하는 Protocol 정보들을 표현하며, 본 C-ITS 환경 모델링의 경우 OBU와 같이 양 기기가 통신하기 위해서는 공통된 Protocol이 존재하여야 한다.
        \end{itemize}
\end{itemize}

또한, RSU 정보를 표현하는 데이터형과 그 예시는 \cref{des_RSU}와 같다.

\begin{table}[ht!]
    \begin{tabular}{c|l|c}
    \hline
    키 이름 & 데이터 타입 & 예시 \\\hline
    \hline
    id & String & RSU\_Streetlamp\\\hline
    Loc\_Road & String & Road\_3 \\\hline
    Con\_Network & String & Network\_C-ITS \\\hline
    Groups & String & RSU\_Indicate \\\hline
    Description & String & Street light \\\hline
    System\_Env & String Array & [``Azure RTOS OS version 2.1.14.1''] \\\hline
    Command & String Array & [``Send mal-information to OBU''] \\\hline
    Protocol & String Array & [``Ethernet'', ``WTP'']\\\hline
    \end{tabular}
    \caption{Description of the Road Side Unit.}
    \label{des_RSU}
\end{table}

\subsubsection{Device}
Device란, RSU 또는 OBU 이외에 C-ITS 환경에서 사용될 수 있는 각종 장치를 의미한다.
Device는 여러 사용자가 공적으로 혹은 사적으로 사용되는 각종 장치를 포괄하며, 특수한 Device의 경우 C-ITS 센터에 접근할 수 있다.
Device 정보 또한, 다양한 정보를 C-ITS 환경 모델링 상에서 표현하기 위해 JSON 포맷을 이용하여 키-값 쌍으로 구성하였다. Device 정보를 표현하는 키 이름과 설명은 다음과 같이 표현된다.

\begin{itemize}
    \item id
        \begin{itemize}
            \item C-ITS 환경 내에서 각 Device를 구분하는 고유 식별자를 의미하는 field를 의미한다.
        \end{itemize}
    \item sub\_id
            \begin{itemize}
            \item 각 Device를 상세 구분하는 식별자를 표현하는 field로, device의 개괄적인 정보를 의미한다.  
        \end{itemize}
    \item Con\_Network
            \begin{itemize}
            \item Device가 연결된 네트워크 식별자를 의미하는 field를 의미한다.
        \end{itemize}
    \item Groups
            \begin{itemize}
            \item Device 기기가 속한 그룹을 표현하는 field로, 그룹 별 사용 목적이 다르므로 Device는 그룹에 따라 사용 용도를 유추할 수 있다. 
        \end{itemize}
    \item Description
            \begin{itemize}
            \item Device의 상세 정보를 표현하는 field로, 각 device에 대한 설명 및 정보를 나타낸다. 
        \end{itemize}
    \item System\_Env
            \begin{itemize}
            \item Device가 가진 환경 정보 (\textit{e.g.}, 시스템) field를 의미한다.
        \end{itemize}
    \item Command
            \begin{itemize}
            \item Device에서 발생된 Command를 의미하며, 내부에서 발생한 명령을 서술식으로 표현한다. 
        \end{itemize}
    \item Protocol
            \begin{itemize}
            \item Device가 사용하는 Protocol 정보들을 표현하며, 본 C-ITS 환경 모델링의 경우 OBU, RSU와 같이 양 기기가 통신하기 위해서는 공통된 Protocol이 존재하여야 한다. 
        \end{itemize}
\end{itemize}

\begin{table}[ht]
    \begin{tabular}{c|l|c}
    \hline
    이름 & 데이터 타입 & 예시 \\\hline
    \hline
    id & String & Attacker\_Device\\\hline
    sub\_id & String & Attacker\_PC \\\hline
    Con\_Network & String & Network\_Malware \\\hline
    Groups & String & Attacker \\\hline
    Description & String & 공격을 수행하는 attacker pc \\\hline
    System\_Env & String Array & [``Windows 10'', ``Virtual Machine''] \\\hline
    Command & String Array & [``Virtual OBU Setting in Device''] \\\hline
    Protocol & String Array & [``Ethernet''] \\\hline
    \end{tabular}
    \caption{`evice.}
\end{table}

\section{공격 시나리오 모델링}
\label{attack-scenario-modeling}
\subsection{소개}
공격 시나리오 모델링은 C-ITS 환경에서 발생할 수 있는 공격 상황에 대한 시나리오를 작성하고 도식화하여 공격이 발생할 수 있는 경로를 사전에 확인하는 과정이다.
본 기술 문서에서는 시나리오를 정의하기 위해 외부 프레임워크를 참조하였으며, 공격 시나리오의 각 스텝에 따른 상태의 변화를 표현하기 위해 필요한 요소를 정의한다. 

\subsection{구성}
본 기술 문서에서 공격 시나리오는 아래의 요소를 사용하여 구성한다.

\begin{multicols}{2}
    \begin{itemize}
        \item 공격 정보
        \item Property
        \item 취약점 정보
        \item Condition
    \end{itemize}
\end{multicols}

공격 정보는 MITRE ATT\&CK Framework를 기반으로 한다. 해당 Framework에 기술되어 있는 ATT\&CK Tactics와 ATT\&CK Techniques를 활용한다.
취약점 정보는 MITRE Coporation의 CVE 취약점을 기반으로 한다.
Property는 다양한 연구\cite{abdrabou2010probabilistic, biswal2014board, fan2015road, vector, petracca2013board, xue2017roadside, yoo2020implementation}를 참조하여 작성하였으며, Property에 대한 자세한 설명은 후술한다.
마찬가지로 Condition은 다양한 연구를 참조하여 작성하였으며, 자세한 설명은 후술한다.

\subsection{공격 정보}
\subsubsection{개요}
본 기술 문서에서 공격 정보는 MITRE ATT\&CK Framework\cite{mitre}를 기반으로 한다. 
본 기술 문서에서 설명하는 Attack Scenario는 ATT\&CK for Enterprise와 ATT\&CK for Mobile을 활용하여 공격자의 행위를 모델링하였다.
아래는 본 기술 문서에서 설명하는 Attack Scenario를 모델링하는데 참조한 ATT\&CK Framework의 Tactics와 Techniques를 설명한다. 

\subsubsection{구성요소}
본 절은 본 기술 문서에서 ATT\&CK for Enterprise와 ATT\&CK for Mobile로부터 참조한 Tactics와 Techniques를 설명한다.

\paragraph{Tactics}
ATT\&CK Tactics는 공격하려는 대상에 대한 공격자의 행동을 의미하며, \cref{table_attack_tactics}와 같다.

\begin{table}[ht!]
\begin{center}
{\small
\resizebox{\textwidth}{!}{
\begin{tabular}{ c|c } 
\hline
\textbf{ATT\&CK Tactic} & \textbf{Description} \\ [0.5ex] 
\hline\hline
Reconnaissance & 향후 계획에 사용 가능한 정보 수집 \\ [1ex]
\hline
Resource Development & 계획을 지원하는데 사용 가능한 리소스 구축 \\ [1ex]
\hline
Initial Access & 공격자의 네트워크 침입 \\ [1ex]
\hline
Execution & 공격자의 악성 코드 실행 \\ [1ex]
\hline
Privilege Escalation & 공격자의 더 높은 권한 획득 \\ [1ex]
\hline
Credential Access & 리소스에 접근할 수 있는 계정 이름, 암호 혹은 기타 비밀을 도용 \\ [1ex]
\hline
Discovery & 공격자의 목표 시스템에 대한 환경 파악 \\ [1ex]
\hline
Lateral Movement & 공격자의 피해자 환경 통과 \\ [1ex]
\hline
Collection & 공격자의 목표로 하는 데이터 수집 \\ [1ex]
\hline
Exfiltration & 공격자의 데이터 훔치기 \\ [1ex]
\hline
Impact & 공격자의 장치와 데이터를 조작, 중단 혹은 파기 \\ [1ex]
\hline
\end{tabular}
}}
\end{center}
\caption{Description of the ATT\&CK Tactics.}
\label{table_attack_tactics}
\end{table}

\paragraph{Techniques}
ATT\&CK Techniques는 공격자가 tactics를 달성하기 위한 방안을 의미하며, 상세 내용은 \cref{tactics_tequniques_table}와 같다.

\begin{table}[ht!]
{\scriptsize
\resizebox{\textwidth}{!}{
\begin{tabular}{ c|c|c } 
\hline
\textbf{ATT\&CK Tactic} & \textbf{ATT\&CK Technique} & \textbf{Description} \\ [1ex] 
\hline
Reconnaissance & Gather Victim Identity Information & 목표 대상에 대한 사용할 수 있는 피해자의 신원 정보 수집 \\ [1ex]
& Gather Victim Org Information & 목표 대상에 대한 사용할 수 있는 피해자의 조직 정보 수집 \\ [1ex]
\hline
Resource Development & Establish Accounts & 공격에 사용 가능한 서비스 계정 생성 및 유지 \\ [1ex]
\hline
Initial Access & Phishing & 피해자 시스템에 액세스하기 위해 피싱 메시지 전송 \\ [1ex]
& Supply Chain Compromise & 피해자의 제품 수령 전에 제품 및 전달 메커니즘 조작 \\ [1ex]
& Replication Through Removable Media & USB를 통해 연결된 장치에 맬웨어 악용 혹은 복사 \\ [1ex]
& Hardware Additions & 시스템이나 네트워크에 하드웨어 혹은 기타 장치를 추가 \\ [1ex]
\hline
Execution & Scheduled Task / Job & 작업 예약 기능을 악용한 악성코드 초기 및 반복 실행 \\ [1ex]
& Exploitation for Client Execution & 클라이언트의 소프트웨어 취약점을 악용한 코드 실행 \\ [1ex]
& Command and Scripting Interpreter & 명령, 스크립트 혹은 바이너리 실행을 위한 명령 및 스크립트 인터프리터 남용 \\ [1ex]
\hline
Privilege Escalation & Valid Accounts & 기존 계정의 자격 증명을 획득 후 남용 \\ [1ex]
& Abuse Elevation Control Mechanism & 상승된 권한 제어를 위해 설계된 메커니즘 우회 \\ [1ex]
& Exploitation for Privilege Escalation & 소프트웨어 취약점을 활용한 권한 상승 \\ [1ex]
\hline
Credential Access & Input Capture & 사용자 입력 캡쳐를 통한 정보 수집 혹은 자격 증명 획득 \\ [1ex]
\hline
Discovery & Network Sniffing & 네트워크 트래픽에서 환경에 대한 정보 캡처 \\ [1ex]
\hline
Lateral Movement & Internal Spear Phishing & 추가 정보에 대한 액세스 권한 획득 혹은 조직 내 다른 사용자 악용 \\ [1ex]
& Lateral Tool Transfer & 손상된 환경에서 시스템 간 도구 혹은 기타 파일 전송 \\ [1ex]
\hline
Collection & Automated Collections & 자동화된 기술을 사용하여 내부 데이터 수집 \\ [1ex]
\hline
Exfiltration & Exfiltration Over Other Network Medium & 명령 및 제어 채널이 아닌 다른 네트워크 매체를 통해 데이터 유출 \\ [1ex]
\hline
Impact & Service Stop & 합법적인 사용자가 서비스 사용이 불가하도록 서비스 중지 혹은 비활성화 \\ [1ex]
& Data Manipulation & 외부 결과 조작을 위해 데이터의 삽입, 삭제 혹은 변경 \\ [1ex]
\hline
\end{tabular}
}}
\caption{Description of the ATT\&CK Techniques included in each ATT\&CK Tactic.}
\label{tactics_tequniques_table}
\end{table}

\subsection{취약점 정보}
\subsubsection{CVE}
CVE란 널리 공개된 소프트웨어 취약점을 정리한 표준 코드를 의미하며, 각 취약점은 모두 고유한 번호를 부여받는다\cite{cve}.
CVE 정보의 주요 데이터 필드는 \cref{table_cve}과 같다.

\begin{table}[h]
\begin{center}
\begin{tabular}{l|l}
\hline
이름                  & 설명           \\\hline
CVE-ID              & CVE 고유 식별자   \\\hline
Description         & 취약점 정보       \\\hline
References          & 참고 자료        \\\hline
Assigning CNA       & 부여 기관        \\\hline
Date Record Created & CVE ID 부여 시기
\\\hline
\end{tabular}
\caption{Description of CVE data fields}
\end{center}
\label{table_cve}
\end{table}

\subsubsection{공격 대상에 따른 CVE 분류}
공격 대상에 따른 CVE 분류는 \cref{table_cve_description}과 같다.
\begin{table}[!ht]
\resizebox{\textwidth}{!}{%
\begin{tabular}{l|l|l|l|l}
\hline
Target & CVE Name & CVE Description & Target Information\\ \hline
\multirow{17}{*}{Device} & CVE-2009-3274 & 다운로드 창에서 선택한 파일에 대해 예측 가능한 /tmp 경로 이름을 사용 & Mozila Firefox 3.6a1, 3.5.3, 3.5.2\\
                        & CVE-2011-1327 & Trend Micro Internet Security 의 키입력 암호화 기능은 암호를 완전히 암호화하지 않아 키로거로 활용 가능한 취약점 & Trend Micro Internet Security 2009\\
                        & CVE-2017-17556 & Synaptics TouchPad 드라이버의 디버그 도구를 활용 키보드 스캔 코드 정보를 획득 가능한 취약점 & Windows(Labtop)\\
                        & CVE-2019-5965 & Joruri Mail 2.1.4 및 이전 버전의 공개 redirection 취약점 & Joruri Mail Version $\sim$ 2.1.4\\
                        & CVE-2019-0976 & Visual Studio 2012 버전 이후로 자동 설치되어 있는 NuGet 패키지 관리자의 변조 취약점 & Visual Studio 2012 Nuget Package\\
                        & CVE-2019-9705 & Synaptics TouchPad 드라이버의 디버그 도구를 활용 키보드 스캔 코드 정보를 획득 가능한 취약점 & 3.0pl1-133 데비안 패키지 이전\\
                        & CVE-2020-10548 & rConfig 3.9.4 및 이전 버전에서의 php SQL Injection 취약점 & rConfig Software Version $\sim$ 3.9.6\\
                        & CVE-2020-10549 & rConfig 3.9.4 및 이전 버전에서의 php SQL Injection 취약점 & rConfig Software Version $\sim$ 3.9.7\\
                        & CVE-2020-11114 & Bluetooth 장치는 L2CAP 페이로드 길이를 적절히 제한하지 않아 버퍼 오버플로우를 통해 악성코드 실행 & Bluetooth\\
                        & CVE-2021-0365 & `mobile\_log\_d'에 잘못된 입력의 유효성 검사를 통해 명령 주입 & Android Version 10, 11\\
                        & CVE-2021-32606 & net/can/isotp.c의 isotp\_setsocopt에 대한 UAF (use-after-free) 취약점 & Linux Kernel 5.11 $\sim$ 5.12.2\\
                        & CVE-2021-39982 & Phone Manager 애플리케이션의 권한 관리 취약점 & Mobile Phone Manager Application\\
                        & CVE-2022-28195 & NVIDIA Jetson Linux Driver 패키지의 Cboot ext4\_read\_file 기능 취약점 & NVIDIA Jetson Linux Driver\\
                        & CVE-2022-28779 & 버전 1.7.50 이전 버전의 Samsung Android USB Driver Windows Installer 프로그램의 취약점 & Samsung Android USB Driver Windows Installer\\
                        & CVE-2022-29581 & Linux 커널의 net/sched의 참조 카운트 취약점의 부적절한 업데이트로 인한 취약점 & Linux Kernel 4.14 $\sim$ 5.18\\
                        & CVE-2022-30155 & Windows Kernel의 DoS 취약점 & Windows 10\\
                        & CVE-2022-31127 & Next.js 서버에서 사용하는 인증 솔루션인 NextAuth.js 취약점 & NextAuth.js $\sim$ 4.9.0\\\hline
\multirow{5}{*}{OBU} & CVE-2018-9318 & Telemetics Control Unit에서 Cellular network를 통해 remote attack을 수행하는 BMW 취약점 & BMW I, X, 3, 5, 7 시리즈\\
                    & CVE-2018-9322 & USB 또는 OBD-II interface를 통해 접근한 후 firmware update를 통해 code-signing 보호를 우회할 수 있는 취약점 & BMW I, X, 3, 5, 7 시리즈\\
                    & CVE-2020-10546 & rConfig 3.9.4 및 이전 버전에서의 php SQL Injection 취약점 & 3.0pl1-133 데비안 패키지 이전\\
                    & CVE-2020-10547 & rConfig 3.9.4 및 이전 버전에서의 php SQL Injection 취약점 & rConfig Software Version $\sim$ 3.9.5\\
                    & CVE-2020-25671 & Linux Kernel의 `Fllcp\_sock\_connect()'의 취약점 & Linux\\\hline
\multirow{4}{*}{RSU} & CVE-2013-0217 & 3.7.8 이전 Linux kernel의 Xen netback 기능에 존재하는 netback.c의 메모리 누수 취약점 & Linux Kernel $\sim$ 3.7.8\\
                    & CVE-2018-9311 & Telemetics Control Unit 권한 상승 취약점 & Linux Yocto\\
                    & CVE-2020-24588 & IEEE 802.11 표준에서 일반 텍스트 QoS 헤더 필드의 A-MSDU 플래그의 인증 불필요 취약점 & Network\\
                    & CVE-2022-23120 & Linux Version 20 이하용 Workload Security Agent의 취약점 & Linux Version $\sim$20\\\cline{1-4} 
\end{tabular}}
\caption{Description of the CVE classification by target.}
\label{table_cve_description}
\end{table}

\subsection{공격 스탭}
공격 스탭은 Conditions와 Command, Property로 구성되어 있으며, 상세 내용은 다음과 같다.

\subsubsection{Conditions}
Condition이란, Attack Scenario 상에서의 상태 변화를 표현한 것으로, 각 Attack Step에서 필요한 환경 조건을 정의한다. 이는 Pre-condition, Post-condition으로 정의되며, 이 때 요구되는 상태 변화를 표현한다.

\paragraph{Pre-Condition}
Pre-Condition은 Attack Step이 실행되기 전 Attack Step이 실행되기 위해 공격자와 공격 대상의 기기가 만족해야하는 조건을 의미한다.

\paragraph{Post-Condition}
Post-Condition은 Attack Step이 실행된 후 변화한 공격자와 공격 대상의 기기에 대한 상태를 의미한다.

\subsubsection{Command}
Command는 공격자가 수행한 행위를 표현한 것으로, 각 Attack Step에서 발생한 실제 행위를 서술한다.

\subsubsection{Property}
Property는 Attack Step 에 따라 변화하는 상태를 표현한 것으로, 각 Attack step을 구성하기 위해 필요한 환경 정보를 정의한다. Property는 각 구성요소마다 다르며, 이와 관련한 사항은 3.2에서 확인할 수 있다. Pre-condition, Post-condition을 정의할 때 요구되는 상태를 표현한다.

\subsection{예시}
\subsubsection{개요}
본 기술문서에서는 C-ITS 환경에서 발생할 수 있는 공격 상황에 대한 Attack Scenario를 예로 들어 설명한다.
\begin{itemize}
    \item Scenario Overview : 공격자는 OBU를 통해 전달되는 데이터를 변조함으로써 도로교통 혼잡 발생
\end{itemize}

예시로 제안된 시나리오에서 공격자는 차량의 USB 취약점을 이용해 USB 포트를 통해 차량 OBU로 접속한다. 이 때, 차량은 BMW 차량의 특정 시리즈로 제한된다. 이어서 공격자는 접속한 OBU를 통해 공격 목표로 선정한 RSU에 통신을 연결한다. 이후 통신에 사용되는 패킷을 캡쳐하여 분석 후 RSU에 패킷을 변조하여 전송한다. 결과적으로 이는 RSU가 변조된 패킷 정보를 바탕으로 잘못된 정보를 타 RSU 혹은 타 OBU에 전송하는 것으로 잘못된 결과를 이끌어내도록 한다.

\subsubsection{공격자 OBU 환경}
\begin{itemize}
    \item Linux Yocto
    \item 128MB NAND DDR3 RAM
    \item IEEE 802.11p, IEEE 1609.2/3/4, DSRC 기능 지원
\end{itemize}

\subsubsection{공격 대상 RSU 환경}
\begin{itemize}
    \item IEEE 802.11 표준 지원
    \item Azure RTOS OS version ~ 2.1.1.1
\end{itemize}

\subsubsection{공격 시나리오}
\begin{figure}[ht!]
    \centering
    \includegraphics[width=1\linewidth]{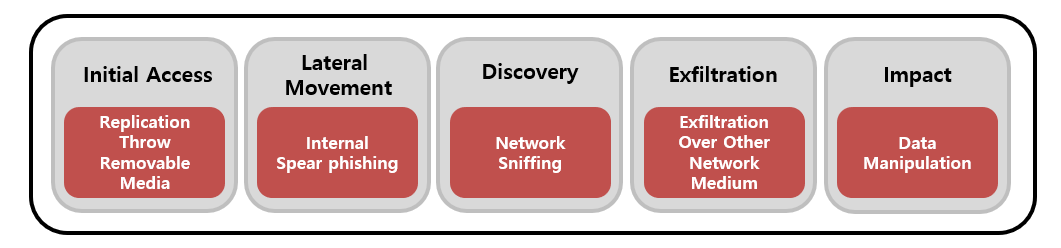}
    \caption{Attack Scenario TTPs.}
    \label{Attack_Scenario_TTPs}
\end{figure}
\begin{enumerate}
    \item Step 1
    
    \begin{figure}[ht!]
    \centering
    \includegraphics[width=0.8\linewidth]{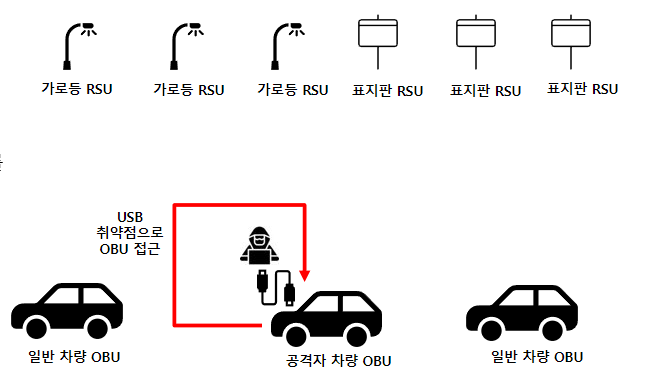}
    \caption{Attack Scenario Step 1.}
    \label{Attack_Scenario_Step1}
    \end{figure}
    
    공격자는 \cref{Attack_Scenario_Step1}과 같이 USB 취약점을 이용, 차량 내 USB 포트를 통해 On Board Unit (OBU)에 접근
    
    \begin{enumerate}
        \item MITRE ATT\&CK Framework 분류
        \begin{itemize}
            \item Initial Access (Tactics) : Replication Through Removable Media
        \end{itemize}
        \item CVE Vulnerability
        \begin{itemize}
            \item CVE-2018-9322
            \begin{itemize}
                \item USB 또는 OBD-II interface를 통해 접근한 후 firmware update를 통해 code-signing 보호를 우회할 수 있는 BMW I, X, 3, 5, 7 시리즈 취약점
                \item 공격자가 root shell 획득 가능
            \end{itemize}
        \end{itemize}
        \item Attack Step 1 Property
        \begin{itemize}
            \item Pre-Condition
                \begin{itemize}
                    \item Attacker는 공격자 차량에 물리적으로 접근할 수 있을 것 (Condition)
                    \item 공격자 차량은 BMW I, X, 3, 5, 7 시리즈 중 하나일 것 (Property)
                    \item 공격자 차량은 USB 포트를 통해 접근할 수 있을 것 (Property)
                \end{itemize}
            \item Command
                \begin{itemize}
                    \item``CVE-2018-9322'' 취약점을 활용, Firmware update를 통해 code-signing 보호 우회
                \end{itemize}
            \item Post-Condition
                \begin{itemize}
                    \item 공격자가 root 권한 획득
                \end{itemize}
        \end{itemize}
    \end{enumerate}
    \item Step 2
    
    \begin{figure}[ht!]
    \centering
    \includegraphics[width=0.8\linewidth]{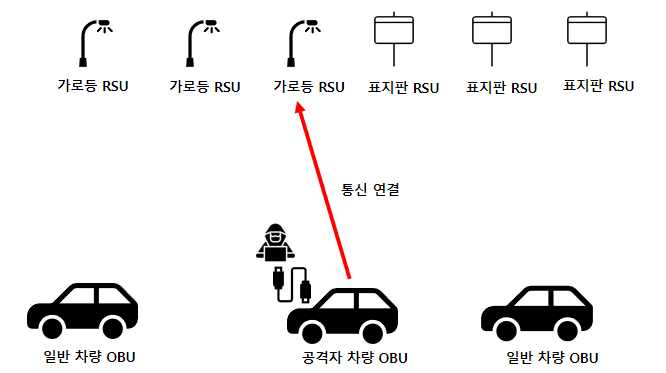}
    \caption{Attack Scenario Step 2.}
    \label{Attack_Scenario_Step2}
    \end{figure}
    
    공격자는 \cref{Attack_Scenario_Step2}와 같이 접근된 차량 내부 OBU와 공격하려는 RSU 간 통신 연결 수행
    
    \begin{enumerate}
        \item MITRE ATT\&CK Framework 분류
            \begin{itemize}
                \item Lateral Movement (Tactics) : Internal Spear phishing
            \end{itemize}
        \item CVE Vulnerability
        \begin{itemize}
            \item CVE-2018-9311
            \begin{itemize}
                \item Telemetics Control Unit 권한 상승 취약점
                \item 공격자가 Telemetics Control Unit 접근 가능
            \end{itemize}
        \end{itemize}
        \item Attack Step 2 Property
        \begin{itemize}
            \item Pre-Condition
                \begin{itemize}
                    \item 공격자는 공격자 차량의 Telemetics Control Unit에 접근 가능할 것 (Condition)
                    \item 공격자 차량과 가로등 RSU는 서로 통신할 수 있을 것 (Condition)
                    \item 공격자 차량과 가로등 RSU 간 통신할 수 있는 공통된 Protocol Property를 가질 것 (Property)
                    \item 공격자 차량의 Device Property 중 Telemetics Control Unit이 포함되어 있을 것 (Property)
                \end{itemize}
            \item Command
                \begin{itemize}
                    \item ``CVE-2018-9311'' 취약점을 활용, 공격자 차량의 Telemetics Control Unit에 remote 공격 수행
                \end{itemize}
            \item Post-Condition
                \begin{itemize}
                    \item 공격자가 Telemetics Control Unit에 대한 접근 및 제어 가능
                \end{itemize}
        \end{itemize}
    \end{enumerate}
    \item Step 3
    
    \begin{figure}[ht!]
    \centering
    \includegraphics[width=0.8\linewidth]{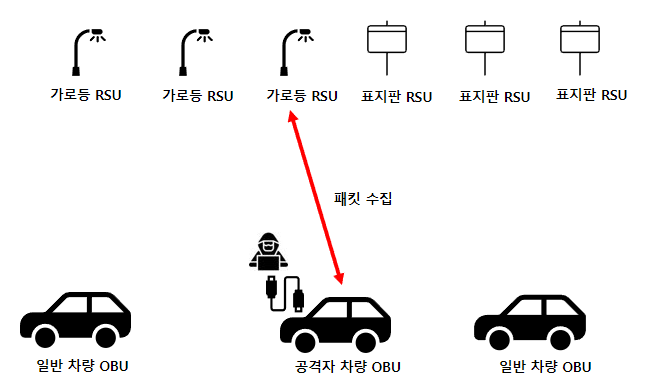}
    \caption{Attack Scenario Step 3.}
    \label{Attack_Scenario_Step3}
    \end{figure}
    
    공격자는 \cref{Attack_Scenario_Step3}과 같이 연결된 통신에서 변조에 필요한 패킷 수집
    
    \begin{enumerate}
        \item MITRE ATT\&CK Framework 분류
         \begin{itemize}
            \item Discovery (Tactics) : Network Sniffing
        \end{itemize}
        \item CVE Vulnerability
        \begin{itemize}
            \item CVE-2018-9318
            \begin{itemize}
                \item Telemetics Control Unit에서 Cellular network를 통해 remote attack을 수행하는 BMW 취약점
                \item 공격자가 원격으로 공격 수행 가능
            \end{itemize}
        \end{itemize}
        \item Attack Step 3 Property
        \begin{itemize}
            \item Pre-Condition
                \begin{itemize}
                    \item 공격자는 공격자 차량과 연결되어 있을 것 (Condition)
                    \item 공격자 차량과 가로등 RSU는 서로 공통된 Protocol Property를 사용해 통신하고 있을 것 (Property)
                \end{itemize}
            \item Command
                \begin{itemize}
                    \item ``CVE-2018-9318'' 취약점을 활용, Telemetics Control Unit에 remote attack 수행
                \end{itemize}
            \item Post-Condition
                \begin{itemize}
                    \item 공격자가 원격으로 변조에 필요한 패킷 수집
                \end{itemize}
        \end{itemize}
    \end{enumerate}
    \item Step 4
    
    \begin{figure}[ht!]
    \centering
    \includegraphics[width=0.8\linewidth]{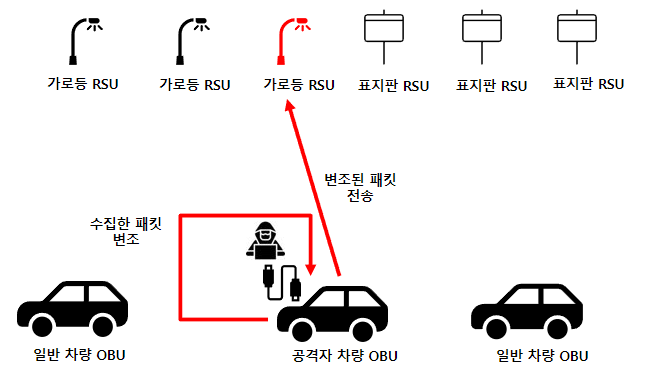}
    \caption{Attack Scenario Step 4.}
    \label{Attack_Scenario_Step4}
    \end{figure}
    
    공격자는 \cref{Attack_Scenario_Step4}와 같이 수집된 패킷 정보를 바탕으로 패킷을 변조하고, 공격하려는 RSU에 임의적으로 변조된 패킷 전송
    
    \begin{enumerate}
        \item MITRE ATT\&CK Framework 분류
            \begin{itemize}
                \item Exfiltration (Tactics) : Exfiltration Over Other Network Medium
            \end{itemize}
        \item CVE Vulnerability
            \begin{itemize}
                \item CVE-2018-9318
                    \begin{itemize}
                        \item Telemetics Control Unit에서 Cellular network를 통해 remote attack을 수행하는 BMW 취약점
                        \item 공격자가 원격으로 공격 수행 가능
                    \end{itemize}
            \end{itemize}
        \item Attack Step 4 Property
        \begin{itemize}
            \item Pre-Condition
                \begin{itemize}
                    \item 공격자는 공격자 차량과 연결되어 있을 것 (Condition)
                    \item 공격자는 공격자 차량의 Telemetics Control Unit에 접근 가능할 것 (Condition)
                    \item 공격자 차량과 가로등 RSU는 서로 송, 수신할 수 있을 것 (Condition)
                    \item 공격자 차량과 가로등 RSU 간 통신할 수 있는 공통된 Protocol Property를 가질 것 (Property)
                \end{itemize}
            \item Command
                \begin{itemize}
                    \item ``CVE-2018-9318'' 취약점을 활용, Telemetics Control Unit을 이용해 remote 공격 수행
                \end{itemize}
            \item Post-Condition
                \begin{itemize}
                    \item 공격자가 원격으로 변조된 패킷 전송
                \end{itemize}
        \end{itemize}
    \end{enumerate}
    \item Step 5
    
    \begin{figure}[ht!]
    \centering
    \includegraphics[width=0.8\linewidth]{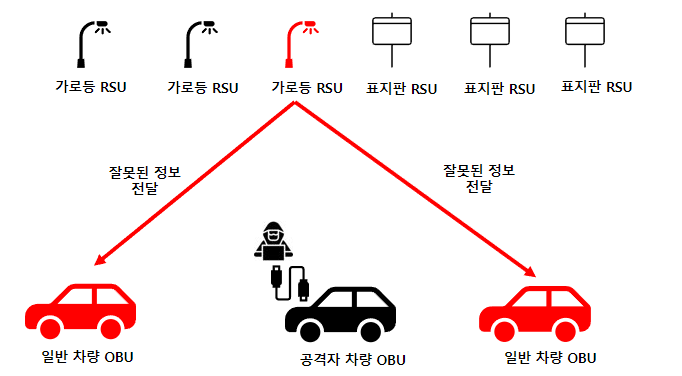}
    \caption{Attack Scenario Step 5.}
    \label{Attack_Scenario_Step5}
    \end{figure}

    \cref{Attack_Scenario_Step5}와 같이 변조된 패킷을 전달받은 RSU는 공격자가 가려는 이동 경로를 공사 중으로 판단하고 다른 차량들에게 전달함
    
    \begin{enumerate}
        \item MITRE ATT\&CK Framework 분류
        \begin{itemize}
            \item Impact (Tactics) : Data Manipulation
        \end{itemize}
        \item Attack Step 5 Property
        \begin{itemize}
            \item Pre-Condition
                \begin{itemize}
                    \item 가로등 RSU와 일반 차량 OBU는 서로 송, 수신할 수 있을 것 (Condition)
                    \item 가로등 RSU와 일반 차량 OBU 간 통신할 수 있는 공통된 Protocol Property를 가질 것 (Property)
                \end{itemize}
            \item Command
                \begin{itemize}
                    \item 가로등 RSU는 변조된 패킷을 바탕으로 도출된 잘못된 정보를 일반 차량 OBU에 전달함
                \end{itemize}
            \item Post-Condition
                \begin{itemize}
                    \item 일반 차량은 잘못된 정보를 바탕으로 공격자에게 이득이 되는 행동을 수행함
                \end{itemize}
        \end{itemize}
    \end{enumerate}
\end{enumerate}

\bibliographystyle{plain}
\bibliography{main}

\begin{figure}[b] 
\centering
\subfloat{
\includegraphics[width=0.35\linewidth]{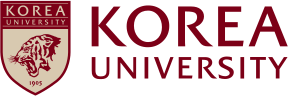}
}
\centering
\subfloat{
\includegraphics[width=0.35\linewidth]{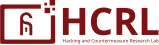}
}

%
\end{figure}

\end{document}